\documentclass{kluwer}
\usepackage{amsmath}
\usepackage{amsthm}
\begin{document}
\begin{opening}
\title{What is the question that MaxEnt answers?} 
\subtitle{A probabilistic interpretation}
\author{Marian \surname{Grend\'ar}}
\institute{ Institute of Measurement Science, Slovak Academy of
Sciences, D\'u\-brav\-sk\'a cesta~9, 842 19 Bratislava, Slovakia
\email{umergren@savba.sk}}
\author{Mari\'an \surname{Grend\'ar}}%
\institute{ Railways of Slovak Republic, DDC, Klemensova 8, 813 61 Bratislava,
Slovakia\email{grendar.marian@zsr.sk}}
\classification{MSC 1991 codes}{Primary 60C05; Secondary 60F99, 62A15, 62A99}
\abbreviations{%
\abbrev{MaxEnt}{Maximum Entropy method},
\abbrev{REM}{Relative Entropy Maximization},
\abbrev{MaxProb}{Maximum Probability method},
\abbrev{ExpOc}{Expected Occurrence vector method}
}
\begin{abstract}
The Boltzmann-Wallis-Jaynes' multiplicity argument is taken up and
elaborated. MaxEnt is proved and demonstrated to be just an
asymptotic case of
looking for such a vector of absolute frequencies in
a feasible set, which has maximal probability of being generated by a
uniform prior generator/pmf.
\end{abstract}
\end{opening}


\newtheorem{thm}{Theorem}
\newtheorem*{JW}{Wallis--Jaynes Theorem}
\newtheorem{lem}{Lemma}
\newtheorem*{cor}{Corollary}
\newtheorem{qtn}{Question}

\theoremstyle{definition}
\newtheorem{defn}{Definition}
\newtheorem{setup}{Setup}
\newtheorem*{ex}{Example}

\theoremstyle{remark}
\newtheorem*{note}{Note}
\newtheorem*{com}{Comments}

\newcommand{\ud}{\mathrm{d}}
\numberwithin{equation}{section}

\section{Introduction}

Shannon's entropy maximization (MaxEnt) \index{maximum entropy}
interpreted by its proponent, E.T. Jaynes,  as a 'method for
inference  from incomplete information' \cite{Jaynes91}, has a
debatable relationship to the Bayesian method (see
\cite{Jaynes88}, \cite{Zellner88}, \cite{Teddy87}, \cite{Uffink},
\cite{GJM}), and to the Maximum Likelihood method (see
\cite{Jaynes82}, \cite{Golan}, \cite{ali2}, \cite{my99}). MaxEnt has also been
presented as a method for assigning prior probabilities (see
\cite{Jaynes98}), (\cite{Teddy87}, \cite{Uffink}
for a critique), and as an extension of the principle of
insufficient reason. On the theoretical level MaxEnt, as a
statistical method, is in dispute for decades.

MaxEnt has been successfully applied to linear inverse
problems \index{inverse problem} without noise that arise in many branches of science.
Its scope was  substantially extended to the inverse problems with noise
by \cite{GJM}, and further by \cite{AJ},
\cite{MJM}, who have put  MaxEnt into a context of
extremum estimation and inference methods.

Theoretical disputes on the status of MaxEnt among
other statistical methods generally ignore
a fundamental question: 'What is the question MaxEnt answers?'. Two other
problems involving 'What kind of constraints are relevant to bind the entropy
maximization?' and 'What if information on amount of data is also
available?' considerations were left on margin of interest both by
Jaynes and opponents of MaxEnt.

In this article we address three elementary problems of the MaxEnt method.
These problems involve  lack of satisfactory probabilistic rationale, ad-hockery regarding constraints to bind MaxEnt
and impossibility of the method to take an information on amount of data into
consideration.

Concerning the problem of rationale, we shall prove and
demonstrate that  MaxEnt is a special,  {\it asymptotic} case of more
general and self-evident 'principle/method' -- {\it MaxProb\/} -- that focuses on
looking for a vector of absolute frequencies (occurrences) in
a feasible set of vectors, which has maximal probability of being
generated by a prior generator pmf. MaxEnt is also a special case of MaxProb in the
sense of assuming a specific uniform pmf prior generator. In fact, the work
provides probabilistic rationale of a more general, relative entropy
maximization (REM) method. REM is also commonly known as
$I$-divergence minimization, or Kullback-Leibler directed distance minimization.
As a by-product of providing the MaxProb rationale of MaxEnt, the remaining
two problems concerning the issues of relevant constraints and the known-sample-size are
also resolved.

The article is organized as follows: Section 2 reviews an
up-to-date status of work on the three above mentioned problems.
Section 3 introduces MaxProb and states the main Theorem 1
together with an example illustrating the point. Section 4
summarizes briefly consequences of Theorem 1 for MaxEnt. Proof of
the Theorem is given in Appendix A. Appendix B  contains a bonus.

\section{Three MaxEnt problems}

We start with a brief review of three problems pertaining to MaxEnt \index{maximum entropy}.

MaxEnt, in the realm of statistics, lacks a proper rationale.
Moment based probability distribution recovered through
Shannon's entropy maximization have been
characterized as the smooth\-est, the flattest ,
the most uniform given a constraints, the least prejudiced,
the one that maximizes our ignorance while including
the available statistical data. These adjectives usually serve
as the main rationale for employing Shannon's entropy criterion. Rarely,
two justifications Jaynes had developed are recalled. The first
one, Wallis' multiplicity argument, is presented in Chapter 11 of
\cite{Jaynes98}. We give it form of theorem and highlight the
central role played by {\it Jaynes' limiting process\/}.

\begin{JW}
Let $\mathbf{N}$ be a discrete multivariate $m$-dimensional random variable
from multinomial distribution
$$
P({\mathbf{N}} = {\mathbf{n}}) = \pi({\mathbf{n}}|{\mathbf{q}}) = \frac{n!}{n_1!\,n_2!\dots n_m!}
\prod_{i=1}^m q_i^{n_i}
$$
where $\sum_{i=1}^m n_i = n$, and ${\mathbf{q}} = [1/m,\  1/m,\  \dots,
1/m]'$ is uniform.

Then under {\rm Jaynes' limiting process}

1) $n\rightarrow\infty$

2) $n_i\rightarrow\infty$ for $i=1, 2, \dots, m$

3) $n_i/n\rightarrow p_i$, where $p_i$ is a constant,  for $i=1, 2, \dots,
m$,

holds

$$
\frac{\ln \pi({\mathbf{n}}|{\mathbf{q})}}{n} \rightarrow H({\mathbf{p}})
$$
where $H({\mathbf{p}}) = - \sum_{i=1}^m p_i \ln p_i$ is Shannon's
entropy (up to an additive constant, $-\ln m$).
\end{JW}

\begin{proof}
Due to Stirling's formula
(there is a typo at the Stirling's approximation (see formula
(11-17)) in the Jaynes' book)
$$
\ln n! = n \ln n - n + \ln(\sqrt{2\pi n}) + 1/(12n) +
{\mathcal{O}}(1/n^2)
$$
holds
\begin{equation*}
\begin{split}
\ln \pi({\mathbf{n}}|{\mathbf{q}}) = n \ln n - n + \ln(\sqrt{2\pi n}) +
1/(12n) \\
- ( \sum_{i=1}^m n_i \ln n_i - \sum_{i=1}^m n_i + \sum_{i=1}^m \ln(\sqrt{2\pi n_i}) + \sum_{i=1}^m
1/(12n_i) ) \\
+ \sum_{i=1}^m n_i \ln q_i + {\mathcal{O}}(1/n^2)
\end{split}
\end{equation*}
Taking into account that $\sum_{i=1}^m n_i = n$, and the first two
assumptions of the Jaynes' limiting process give
$$
\frac{\ln \pi({\mathbf{n}}|{\mathbf{q}})}{n} \rightarrow \ln n -
(1/n)\sum_{i=1}^m n_i \ln n_i - \ln m
$$
where the RHS is $ - \sum_{i=1}^m \frac{n_i}{n} \ln
\frac{n_i}{n} - \ln m$, which leads thanks to the third assumption of the
limiting process to the claim of the Theorem.
\end{proof}

In order to provide a rationale for Shannon's entropy
maximization one has to investigate each particular set of
constraints binding maximization of $\pi(\mathbf{n}|\mathbf{q})$
to determine whether the constraints induce just the Jaynes'
limiting process. Yet it has not been done, and it seems to be not the case, even for the simplest,
moment consistency constraints (mcc), traditionally accompanying the entropy
maximization. For, even in the case of mcc, there is for any $n$ at least one vector
$\mathbf{n}$ in conflict with the second requirement of Jaynes'
limiting process.

Thus, Wallis' argument, rather than offering explanation of MaxEnt,
presents yet another (and interesting) partial limit
 of multinomial distribution, in addition to the well-known DeMoivre-Laplace and the Poisson one. It
could be better called 'Wallis-Jaynes local limit theorem'.

The second  rationale, Jaynes proposed has been
dubbed 'The Entropy Concentration Theorem', (see \cite{JaynesECT}, \cite{Jaynes82}, or
\cite{Teddy87}). This rationale rests on the Wallis' argument and  Jaynes' limiting
process and again leads to the question about its
relevance to constraints bound entropy maximization.

It is well known that MaxEnt can not incorporate information
on amount of data (see for instance \cite{UffinkME}), or in other words, MaxEnt
recovers the same pmf regardless of the sample size. This second problem of
MaxEnt, is usually left open, although it directly relates to the next,
more frequently debated, third problem concerning the kind of
constraints relevant to bind the entropy maximization.
According to \cite{Jaynes79}, constraints should represent
testable information, i.e. they should be a test basis concerning the probability
distribution. For instance a set of samples
is not testable, value of, let's say, third moment is.
Such a state of art also reveals lack of internal coherence of
the views on entropy maximization.

Three interconnected questions are 'Just what are we accomplishing when we maximize
entropy?' \cite{Jaynes82}, 'What kind of constraints are allowed to bind the entropy
maximization?' and 'What if information on amount of data is also
available?'. Vagueness of answers to them
has given to rise in context of linear inverse
problems to yet another question: 'Why MaxEnt?'. This question has been addressed by several
axiomatizations (see \cite{SJ}, \cite{TTL}, \cite{Skilling},
\cite{Csiszar}, \cite{PV}, \cite{Tono}) which single out the
Shannon's entropy (or relative entropy) as the only function,
consistent with the axioms.
Though these axiomatizations answer the most pragmatic 'Why
MaxEnt?' question, they leave unresolved the first three
('interpretational') problems.

\section {REM/MaxEnt -- as an asymptotic case of MaxProb}

Let us consider following general setup.

Let ${\mathbf{q}} = [q_1, q_2, \dots, q_m]'$ be a pmf, defined on $m$-element support,
referred to as {\it prior generator\/} \index{prior generator}.

Let ${\mathcal{H}}_n$ be a set of  all vectors $\{{\mathbf{n}}_1,
{\mathbf{n}}_2, \dots, {\mathbf{n}}_J\}$, such that an adding-up constraint $\sum_{i=1}^m n_{ij} =
n$, for $j=1, 2, \dots, J$, is satisfied. ${\mathbf{n}}$ will be referred to as {\it
occurrence vector} \index{occurrence vector}, ${\mathcal{H}}_n$ as {\it occurrence-vector working set}.

Let ${\mathcal{P}}$ be a set of all probability vectors, such that
$\sum_{i=1}^m p_i = 1$.

Then, a simple question can be asked.

\begin{qtn}
What is the most probable
occurrence vector ${\hat{\mathbf n}}$, among occurrence vectors
${\mathbf{n}}$ from the working set ${\mathcal{H}}_n$,
to be generated by the prior generator ${\mathbf{q}}$?
\end{qtn}

The answer to the question is
\begin{equation}
\hat{\mathbf{n}} = \arg \max_{{\mathbf{n}} \in {\mathcal{H}}_n}
\pi({\mathbf{n}}|{\mathbf{q}})
\end{equation}
where
\begin{equation}\label{var}
\pi({\mathbf{n}}|{\mathbf{q}}) = \frac{n!}{n_1!\,n_2!\dots\,n_m!} \prod_{i=1}^m q_i^{n_i}
\end{equation}
is the probability of generating the occurrence vector
${\mathbf{n}}$ by a prior generator ${\mathbf{q}}$.

\begin{defn}
The above setup and Question 1, leading to task (3.1), (3.2) will be referred to as MaxProb.
\end{defn}

The following theorem states the main result on asymptotic equivalence
of MaxProb and REM/MaxEnt. The proof is developed in the Appendix A.

\begin{thm}
Let ${\mathbf{q}}$ be the prior generator and ${\mathcal{H}}_n$ be the working
set.
Let $\hat{\mathbf{n}}$ be the most probable occurrence vector from the
working set ${\mathcal{H}}_n$, to be generated by the prior generator ${\mathbf{q}}$.
And let $n\rightarrow\infty$. Then
\begin{equation*}
\frac{\hat{\mathbf{n}}}{n} = \hat{\mathbf{p}}
\end{equation*}
where
\begin{equation*}
\hat{\mathbf{p}} = \arg \max_{{\mathbf{p}} \in
{\mathcal{P}}}  H({\mathbf{p}}, {\mathbf{q}})
\end{equation*}
and
\begin{equation*}
H({\mathbf{p}}, {\mathbf{q}}) = - \sum_{i=1}^m p_i
\ln \left(\frac{p_i}{q_i}\right)
\end{equation*}
is the relative entropy of probability vector ${\mathbf{p}}$ on generator ${\mathbf{q}}$.
\end{thm}

\begin{cor} If also some other {\rm differentiable\/} constraint
$F(\frac{{\mathbf{n}}}{n}) = 0$ is employed to form the working set ${\mathcal{H}}_n$, and
a corresponding constraint $F({\mathbf{p}}) = 0$ is added to  the relative
entropy maximization, the claim of Theorem 1 remains valid.
\end{cor}

\begin{note}
If the prior generator is uniform, $H({\mathbf{p}}, {\mathbf{q}})$
reduces to Shannon's entropy $H({\mathbf{p}}) = - \sum_{i=1}^m p_i \ln p_i$.
\end{note}

\begin{ex}
Let ${\mathbf{q}} = [0.13\ 0.09\ 0.42\ 0.36]'$. Let ${\mathcal{H}}_n$
consists of  all occurrence vectors $\{{\mathbf{n}}_1, {\mathbf{n}}_2, \dots, {\mathbf{n}}_J\}$
such that
\begin{equation}\label{Hconstr}
\begin{split}
\sum_{i=1}^m n_{ij} &= n \quad\text{for\ \ } j = 1, 2, \dots, J \\
\sum_{i=1}^m n_{ij} x_i &= 3.2 n \quad\text{for\ \ } j = 1, 2, \dots, J
\end{split}
\end{equation}
where ${\mathbf{x}} = [1\ 2\ 3\ 4]'$.

Table 1 shows $\frac{\hat{\mathbf{n}}}{n}$ and $J$, for  $n = 10, 50, 100, 500,
1000$, together with probability vector $\hat{\mathbf{p}}$ maximizing
relative entropy $H({\mathbf{p}}, {\mathbf{q}})$ under constraints
\begin{equation}\label{constr}
\begin{split}
\sum_{i=1}^m p_i &= 1 \\
\sum_{i=1}^m p_i x_i &= 3.2
\end{split}
\end{equation}

Results for uniform prior generator are in the fourth column of the table.

\begin{table}[!ht]
\caption{MaxProb and REM/MaxEnt}
\renewcommand\arraystretch{1.5}
\begin{tabular}{c|c|c|c}
$n$ & $J$ & $\frac{\hat{\mathbf{n}}}{n}|\mathbf{q}$  &  $\frac{\hat{\mathbf{n}}}{n}|$\ uniform prior\\
\hline
10  &    10 & 0.1000 0.0000 0.5000 0.4000  & 0.1000 0.1000 0.3000 0.5000\\
50  &   154 & 0.0800 0.0600 0.4400 0.4200  & 0.0800 0.1400 0.2800 0.5000\\
100 &   574 & 0.0800 0.0700 0.4200 0.4300  & 0.0800 0.1400 0.2800 0.5000\\
500 & 13534 & 0.0820 0.0700 0.4140 0.4340  & 0.0780 0.1460 0.2740 0.5020\\
1000 & 53734& 0.0830 0.0700 0.4110 0.4360  & 0.0790 0.1460 0.2710 0.5040\\
\hline
$\hat{\mathbf{p}}$ & &0.0826 0.0709 0.4103 0.4361 & 0.0788 0.1462 0.2714 0.5037
\end{tabular}
\end{table}
\end{ex}

\section{Concluding remarks}

As a way of summing up we make the following points:

1) The multiplicity argument, as it is left to us by Boltzmann, Wallis,
Jaynes, does not provide  satisfying
rationale for MaxEnt. However it does provide a clue where to search for
it.

2) Theorem 1, as any asymptotic theorem, can be
interpreted in two directions: either moving towards infinity
or moving back to finiteness. The first direction
provides the MaxProb rationale for REM/MaxEnt. The
second direction shows, that proper place of REM/MaxEnt is only in the asymptotic.

3) MaxProb also automatically solves the known-sample-size problem.

4) Proof of Theorem 1 implies, that constraints that bind REM/MaxEnt
should be differentiable.

5) Hidden behind (3.1), (3.2) is an assumption about {\it iid}
sampling. Thus, REM/MaxEnt as an asymptotic form of MaxProb, seem to
be limited to the iid case.

The probabilistic rationale for REM/MaxEnt/$I$-divergence
minimization may also be extended to provide a rationale of $J$-divergence
minimization \index{minimum I and J divergence}, see \cite{GG2000}.

Area of applicability of MaxProb/REM/MaxEnt should be obvious --
wherever Question 1 is reasonable to ask.

Philosophical consequences of Theorem 1 are left to the reader.

\section{Acknowledgements}

It is a pleasure to thank Ariel Caticha, Peter Cheeseman, Ale\v s Gottvald,
George Judge, Teddy Seidenfeld, Alberto Solana and Viktor Witkovsk\'y for
valuable discussions and/or comments on earlier version of this paper.

\newpage

\appendix

\section{Proof of Theorem 1}

\begin{proof}

1) %
\begin{align*}
\max_{{\mathbf{n}}} \ \,& \pi({\mathbf{n}}|{\mathbf{q}}) \\
& \text{subject to} \\
\sum_{i=1}^m n_i &= n \\
\end{align*}

For the purpose of maximization $\pi({\mathbf{n}}|{\mathbf{q}})$ can
be $log$-transformed, into
\begin{equation*}
v({\mathbf{n}}) = \gamma(1+ \sum_{i=1}^m n_i) - \sum_{i=1}^m \gamma(n_i + 1) + l
\end{equation*}
where $\gamma(\cdot) = \ln \Gamma(\cdot)$,
$\Gamma(\cdot)$ is  gamma-function, and $l = \sum_{i=1}^m n_i \ln
q_i$.
Necessary condition for maximum of $\pi({\mathbf{n}}|{\mathbf{q}})$
than is
\begin{equation}\label{diff}
\ud{v({\mathbf{n}})} = \sum_{i=1}^m \left[- \gamma'(n_i + 1) + \ln q_i\right] \ud{n_i} = 0
\end{equation}
since, according to the assumed adding-up constraint, $\sum_{i=1}^m \ud{n_i} = 0$.

First, it will be proved that
\begin{equation}\label{lim:gamma'}
    \lim_{n_i\to\infty}[\gamma'(n_i)-\ln(n_i+k)] = 0
\end{equation}
for any $n_i$, and any $k$.

The first derivative of $\gamma$ can be written in a form of
infinite series (see \cite{Fichten})
\begin{gather*}
    \gamma'(n_i) = g(n_i) - C
\intertext{where}
    g(n_i) = \sum_{j=0}^\infty\left(\frac{1}{j+1}-\frac{1}{j+n_i}\right)\\
    C  = \text{Euler's constant}
\end{gather*}

For $n_i \in {\mathbb{Z}}$, denoted $\hat{n}_i$, the series
$g(n_i)$ reduces into a harmonic series $H$
\begin{equation*}
g(\hat{n}_i) = \sum_{j=1}^{\hat{n}_i - 1} \frac{1}{j} =
H(\hat{n}_i -1)
\end{equation*}
Let $n_i = \hat{n}_i + h$, where $0\le h < 1$. Then
\begin{equation*}
    \frac{1}{j+1}-\frac{1}{j+\hat{n}_i}
    \leq \frac{1}{j+1}-\frac{1}{j+\hat{n}_i + h}
    < \frac{1}{j+1}-\frac{1}{j+\hat{n}_i  + 1},
\end{equation*}
so
\begin{equation*}
    g(\hat{n}_i) =
    \sum_{j=1}^{\hat{n}_i -1} \frac{1}{j}
    \leq g(n_i) <
    \sum_{j=1}^{\hat{n}_i} \frac{1}{j} = g(\hat{n}_i + 1).
\end{equation*}
Since, difference of the major series converges to zero,
\begin{equation*}
    \lim_{n_i\to\infty}[g(\hat{n}_i +1) - g(\hat{n}_i)]
    = \lim_{n_i\to\infty}\frac{1}{\hat{n}_i} = 0
\end{equation*}
also
\begin{equation}\label{lim1}
    \lim_{n_i\to\infty}[g(n_i) - g(\hat{n}_i)] = 0.
\end{equation}

Due to a known property of harmonic series
\begin{equation*}
    \lim_{n_i\to\infty}[H(n_i)-\ln(n_i + k) - C]=0,\ \text{for  any $k$}
\end{equation*}
holds also
\begin{equation}\label{lim2}
    \lim_{n_i\to\infty}[g(\hat{n}_i)-\ln(n_i + k) - C]=0
\end{equation}
thus, adding \eqref{lim1}, \eqref{lim2} up gives
\begin{equation*}
    \lim_{n_i\to\infty}[g(n_i)-\ln(n_i + k) - C]=0,
\end{equation*}
respectively, for the derivative (recall that $\gamma'(n_i) = g(n_i) - C$)
\begin{equation*}\label{lim3}
    \lim_{n_i\to\infty}[\gamma'(n_i)-\ln(n_i + k)]=0,
\end{equation*}
what is just \eqref{lim:gamma'}.

Without loss of generality, for $n\rightarrow\infty$ we can
restrict for sub-space of $n_i \rightarrow\infty$, for $i=1, 2,
\dots, m$, if on the sub-space exists the maximum, so the
conditions of maximum for $n_i \rightarrow \infty$
take, thanks to \eqref{lim:gamma'}, form
\begin{subequations}
\label{final}
\begin{align}
    \sum_{i=1}^m \left[ - \ln{n_i} + \ln{q_i}\right] \ud{n_i} &= 0 \\
    \sum_{i=1}^m n_i &= n
\end{align}
\end{subequations}

Due to Lemma 1, with $k = \frac{1}{n}$,
system \eqref{final} can be transformed into an equivalent one
\begin{subequations}
\label{end}
\begin{align}
    \sum_{i=1}^m \left[ - \ln \frac{n_i}{n} + \ln{q_i}\right] \ud{\frac{n_i}{n}} &= 0 \\
    \sum_{i=1}^m \frac{n_i}{n} &= 1
\end{align}
\end{subequations}

2) %
\begin{align*}
\max_{{\mathbf{p}}}\ \, & H({\mathbf{p}},{\mathbf{q}}) \\
& \text{subject to} \\
\sum_{i=1}^m p_i &= 1 \\
\end{align*}

Thus necessary conditions for maximum of relative entropy $H({\mathbf{p}},
{\mathbf{q}})$, constrained by the respective adding-up constraint
are
\begin{subequations}
\label{ent}
\begin{align}
     \ud{H({\mathbf{p}}, {\mathbf{q}})} &=
      \sum_{i=1}^m [-\ln p_i + \ln q_i]\ud{p_i} = 0 \\
\sum_{i=1}^m p_i &= 1
\end{align}
\end{subequations}

Comparing \eqref{ent} and \eqref{end} completes the proof.
\end{proof}

\begin{note}
Claim of the Corollary of  Theorem 1 is immediately implied by the  proof.
\end{note}

\begin{lem}
$$
\ud{v(k{\mathbf{n}})} = k\, \ud{v({\mathbf{n}})}
$$
if $\sum_{i=1}^m n_i = n$.
\end{lem}
\begin{proof}
$\ud{v(k{\mathbf{n}})} = \sum_{i=1}^m [ - \ln k n_i  + \ln q_i] \ud{k n_i} =
k \sum_{i=1}^m [ - \ln k - \ln n_i  + \ln q_i] \ud{ n_i} =
k \sum_{i=1}^m [ - \ln n_i + \ln q_i] \ud{n_i} = k\,
\ud{v({\mathbf{n}})}$.
\end{proof}

\newpage

\section{Expected Occurrence Vector (ExpOc) and REM/MaxEnt}

Also, a question different than the Question 1 can be asked.

\begin{qtn}
What is the expected
occurrence vector $\bar{\mathbf{n}}$, of occurrence vectors
${\mathbf{n}}$ from the working set ${\mathcal{H}}_n$,
which can be  generated by the prior generator ${\mathbf{q}}$?
\end{qtn}

Specifying the notion of expectation,
\begin{equation}
\bar{\mathbf{n}} = \frac{\sum_{j=1}^J \pi({\mathbf{n}}_j|{\mathbf{q}})
{\mathbf{n}}_j}{\sum_{j=1}^J \pi({\mathbf{n}}_j|{\mathbf{q}})}
\end{equation}

\begin{thm}
Let ${\mathbf{q}}$ be the prior generator and ${\mathcal{H}}_n$ be the working
set.
Let $\bar{\mathbf{n}}$ be the expected occurrence vector of the
working set ${\mathcal{H}}_n$, to be generated by the prior generator ${\mathbf{q}}$.
And let $n\rightarrow\infty$. Then, using notation of Theorem 1,
$$
\frac{\mathbf{\bar n}}{n} = \hat{\mathbf{p}}
$$
\end{thm}

\begin{cor} If also a moment consistency cosntraint
is employed to form the working set ${\mathcal{H}}_n$, and
corresponding constraint  is added to  the relative
entropy maximization, the claim of Theorem 2  remains valid.
\end{cor}

\begin{note}
Generality of Corollary 2 is under both numerical and theoretical
investigations. Theorem 2 and its Corollary, although yet supported only by numerical calculations,
indicates that MaxEnt/REM can be  as well an asymptotic
form of another method/principle -- {\rm ExpOc} -- which concentrates on looking for an
{\rm expected occurrence vector\/} in the feasible set of vectors generated by a
prior generator/pmf.
\end{note}

Following example illustrates the point of Theorem 2/Corollary 2.

\begin{ex}
Assuming the same setup as in the previous Example, expected
occurrence vectors for $n= [10, 50, 100, 1000]$ and generators
${\mathbf{q}} = [0.13\ 0.09\ 0.42\ 0.36]'$, ${\mathbf{q}} = [0.25\ 0.25\ 0.25\ 0.25]'$
are in Table 2.

\begin{table}[!hm]
\caption{ExpOc and REM/MaxEnt}
\renewcommand\arraystretch{1.5}
\begin{tabular}{c|c|c}
$n$ & $\frac{\bar{\mathbf{n}}}{n}|\mathbf{q}$ & $\frac{\bar{\mathbf{n}}}{n}|$\ uniform prior\\
\hline
10   & 0.0721 0.0736 0.4365 0.4178 & 0.0701 0.1510 0.2877 0.4912 \\
50   & 0.0806 0.0714 0.4153 0.4327 & 0.0771 0.1471 0.2745 0.5013 \\
100  & 0.0816 0.0712 0.4128 0.4344 & 0.0779 0.1466 0.2729 0.5025 \\
500  & 0.0824 0.0710 0.4108 0.4358 & 0.0786 0.1463 0.2717 0.5035 \\
1000 & 0.0825 0.0709 0.4106 0.4360 & 0.0787 0.1462 0.2715 0.5036 \\
\hline
$\hat{\mathbf{p}}$ & 0.0826 0.0709 0.4103 0.4361 & 0.0788 0.1462 0.2714 0.5037
\end{tabular}
\end{table}
Note that $\frac{ \bar{\mathbf{n}} }{n}$ converges to  $\hat{\mathbf{p}}$
faster than $\frac{\hat{\mathbf{n}}}{n}$.
\end{ex}

\newpage

\bibliographystyle{amsplain}

\end{document}